**Title:**
Crystallization of Cuprous Oxide Thin Films by Continuous-Wave Laser Diode with Micro-Chevron Laser Beam ($\mu$-CLB)

**Authors:**
Giraldo, B. [†,‡] , Yeh, W. [¥] , Kobayashi, N.P. [†,‡]

**Affiliations:**

[†] *Baskin School of Engineering, Univ. of California Santa Cruz, Santa Cruz, CA 95064, United States*

[‡] *Nanostructured Energy Conversion Technology and Research (NECTAR), Univ. of California Santa Cruz, Santa Cruz, CA 95064, United States*

[¥] *Shimane University, Matsue, Japan*

## Abstract:

Crystallization of thin film materials by exploiting laser induced crystallization has been advancing for the past four decades. This unique thin film technique has been predominantly used in processing thin film materials made of a single chemical element; however, harnessing this technique to extend its use for thin film materials containing multiple chemical elements (e.g., metal oxides) unlocks applications currently not accessible. In this study, laser crystallization of a $Cu_2O$ strip was demonstrated. A continuous-wave laser diode with a micrometer-scale chevron-shaped beam profile – micro chevron laser beam (µ-CLB) – was used to crystallize CuO thin films covered with an amorphous carbon (a-C) cap layer, deposited on fused silica substrates. Electron backscatter diffraction, Raman spectroscopy, photoluminescence spectroscopy, and UV-Vis spectroscopy were used to investigate the crystallinity and optical properties of the $Cu_2O$ thin films revealing their unique characteristics associated with the crystallization process.

## 1. Introduction:

Laser crystallization is an attractive technique for device structures with minimum thermal budgets because of its capability of heating and treating a thin film locally and selectively, minimizing thermal impacts on the substrate in which a thin film is deposited. Crystallization of a thin film driven by thermal energy provided via a heat source highly localized within the thin film offers a substantial advantage in particular when a substrate is made of materials with low glass-transition temperature or low melting temperature as polymers and covalent-network-glasses, minimizing undesirable physical and chemical interactions between the thin film and the substrate. Laser crystallization has a long history, dating back to the early 1980s [1, 2, 3], with a significant emphasis on thin film materials comprised of a single chemical element like Si, used for the development of thin film transistors [2, 3]. Laser crystallization of semiconductor thin films on arbitrary substrates is of great value because conventional methods by which single-crystal thin films are obtained often employ epitaxial growth that requires expensive precursors, complex process control, and costly single-crystal substrates. Laser crystallization was applied to alloy semiconductor thin films containing multiple chemical elements (e.g., group IV compound semiconductors, group III-V compound semiconductors, metal oxide semiconductors) [4, 5, 6, 7, 8, 9, 10]; however, in all these cases, a femtosecond laser or excimer laser is used, nominal lateral size of crystalline domains are on the order of 1 µm. Using the previously mentioned laser types poses considerable challenges in developing laser crystallization processes that are economically sound and provide crystalline domains large enough for fabricating practical devices. Also, structural and chemical integrity of alloy semiconductor thin films after undergoing laser crystallization processes are always arguable to some extent. In this paper, laser crystallization of $Cu_2O$ thin films was demonstrated using continuous-wave (CW) laser diode (LD) with a micrometer-scale chevron-shaped beam profile –micro chevron laser beam (µ-CLB). The crystallization was induced in thin films made of non-single-crystal CuO (cupric oxide) capped with an amorphous carbon (a-C) layer, resulting in the formation of a single-crystal $Cu_2O$ (cuprous oxide) strip with a semi-infinite volume. These crystallized strips exhibited peculiar optical properties reflecting the unique characteristics of the µ-CLB crystallization process.

## 2. Experimental:

A 130 nm thick CuO thin film was deposited on fused silica substrates and subsequently capped with a 10 nm thick a-C layer. The CuO thin film and a-C capping layer were deposited sequentially by radio frequency (RF) and direct current (DC) magnetron sputtering at room temperature, respectively, in a single vacuum chamber without breaking the vacuum. CuO and C sputtering targets with a purity of 99.99 % were used. A specific thickness of 130 nm was chosen for the CuO thin film to obtain sufficient absorption of the µ-CLB at the wavelength of 405 nm. The real and imaginary parts of the CuO thin film refractive index were measured by spectroscopic ellipsometry and determined to be n = 2.37 and k = 1.01 at 405 nm, respectively. The 10 nm thick a-C capping layer was found to be critical to reducing incongruent evaporation during the crystallization. The µ-CLB that provided laser light with a nominal spot size on the order of 10 µm was generated by having the output of a 405 nm wavelength multimode CW LD pass through a one-sided dove prism that converted the original beam into a chevron shape focused on the thin film sample. The thin film sample was mounted on a linearly moving stage that advanced at a speed of 1 mm/s with respect to the fixed position of the µ-CLB with the laser power output set to 79 mW [11]. A semi-infinite crystallized strip region formed with a width comparable to or less than the nominal spot size 10 µm of the µ-CLB, while the length of the strip is only limited by the linear translational motion of the moving stage and can be extended as needed.

## 3. Results and Discussion:

Electron backscatter diffraction (EBSD) analysis was carried out on a crystallized $Cu_2O$ strip in a scanning electron microscope (SEM) to determine its phase and crystallinity. The crystallized copper oxide strip was identified as $Cu_2O$, also known as cuprite, a cubic crystal system with a lattice parameter of 0.425 nm, belonging to the O4h = $P\bar{n}3m$ space group [12]. Shown in Fig. 1(a) and 1(b) are respectively the EBSD crystallographic orientation map in the normal direction (ND) and laser scanning direction (SD) of the crystallized $Cu_2O$ strip, referred to as single-crystal $Cu_2O$ strip (SC-$Cu_2O$ strip) hereafter. Shown in Fig. 1(c) is the inverse pole figure illustrating a color map corresponding to respective crystal orientations. Presented in Fig. 1(d) is a boundary map of random grain boundaries (RGB, 5°~65°) and coincidence site lattice (CSL) boundaries, indicated by black and red lines respectively. Only a few CSL boundaries are found in the SC-$Cu_2O$ strip. RGBs exist at either side of the strip, and no RGBs completely cross the strip, indicating the strip is a continuous single crystal. RGBs exist periodically on the strip and are synchronized with crest regions of the wave like features in Fig.1(f). Segregation of compounds other than $Cu_2O$ might be the reason. The orientation of domains seen in Fig.1(a) and (b) change gradually and continuously along the strip. Fig.1(e) illustrates the posture of the cubic unit cell seen from ND at corresponding positions in Fig.1(a), crystal orientation is rotating while the crystal advances in a positive pitch direction. Similar orientation rotation phenomenon were reported for a crystallized strip made of Si, except that the pitch rotation is in the opposite direction. In Si, orientation rotation was caused by differences in expansion rate of the thin film at the film-air and film-substrate boundaries during solidification [11]. Positive pitch rotation suggests that the density of $Cu_2O$ is higher in its solid phase than in its liquid phase, or that there is desorption of some component taking place at the surface during solidification. Fig. 1 (f) shows a top-view SEM image of the SC-$Cu_2O$ strip. The white arrow indicates the direction in which μ-CLB advanced with respect to the sample. The overall surface of the SC-$Cu_2O$ strip is textured with wave-like features, periodically found every ~4 μm. Expansion along the center of the SC-$Cu_2O$ strip is similar to that of SC-Si strips formed by μ-CLB [11]. This mushrooming of the solid material about the center of the strip suggests agglomeration of $Cu_2O$ film takes place when melting occurs. The smooth region adjacent to the SC-$Cu_2O$ strip shows a region on the original CuO thin film not subjected to the laser crystallization and referred to as non-single-crystal CuO region (NSC-CuO region) henceforth.

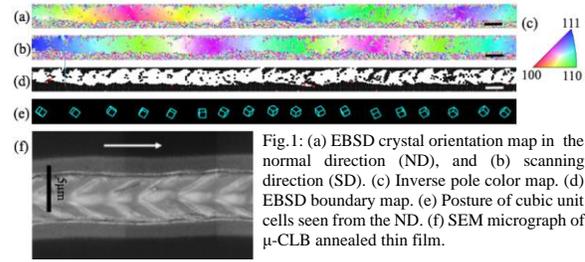

Fig.1: (a) EBSD crystal orientation map in the normal direction (ND), and (b) scanning direction (SD). (c) Inverse pole color map. (d) EBSD boundary map. (e) Posture of cubic unit cells seen from the ND. (f) SEM micrograph of μ-CLB annealed thin film.

Raman spectroscopy analysis was carried out with an excitation wavelength of 514.5 nm to confirm the phase and assess the crystallinity of the SC-$Cu_2O$ strip. Fig.2 shows Raman spectra collected from the SC-$Cu_2O$ strip (red line) and the NSC-CuO region (blue line). The spectrum from the SC-$Cu_2O$ strip shows several phonon modes unique to the crystalline phase of $Cu_2O$, while that from the NSC-CuO region shows no identifiable phonon modes. Characteristic modes (e.g., modes at ~300 cm$^{-1}$ and ~350 cm$^{-1}$) associated with CuO are not seen in Fig. 2, confirming that the SC-$Cu_2O$ strip is predominantly made of $Cu_2O$ [13, 14]. The two phonon modes at 218 cm$^{-1}$ and 436 cm$^{-1}$ represent second and fourth-order overtones, respectively, of the phonon mode at 109 cm$^{-1}$ [15], an inactive Raman mode that is only infrared-allowed in perfect $Cu_2O$ crystal, indicating that the SC-$Cu_2O$ strip bears structural integrity comparable to $Cu_2O$ formed under conditions near thermal equilibrium [16]. The presence of the well-defined second-order overtone centered at 218 cm$^{-1}$ further indicates that the SC-$Cu_2O$ strip has high crystallographic integrity [15]. The mode at 640 cm$^{-1}$ is most likely associated with an allowed L.O. phonon mode [17, 18]. Although complex oxidation kinetics of copper at room temperature resulting in the interplay between the two phases, CuO and $Cu_2O$, would contribute to the Raman analysis [19], the observed phonon modes may largely be attributed to Raman selection rules lifted due to point defects such as Cu vacancies commonly present in p-type $Cu_2O$ [20]. Further study is in progress to assess the dependence of the Raman spectrum on the formation of defects governed by specific parameters set for crystallization by using µ-CLB.

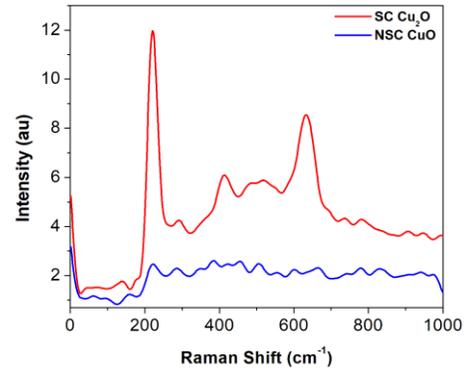

Fig. 2 Raman spectra of non-single crystal (NSC) CuO and single crystal (SC) $Cu_2O$. The SC $Cu_2O$ Raman spectrum was vertically shifted for clarity.

Photoluminescence (PL) spectra of the NSC-CuO region and the SC-$Cu_2O$ strip were collected with a Perkin Elmer luminescence spectrometer equipped with a Xeon lamp. The excitation wavelength used for the PL analysis was 400 nm, and the PL spectra were collected in the spectral range from 1.25 eV to 2.625 eV at room temperature. For the PL measurement, a special coupon with 0.1 mm x 2.0 mm area – strip region – was prepared by crystallizing multiple 10 µm x 2 mm SC-$Cu_2O$ strips spatially separated by a fixed interval of 6 µm. Multiple strips were used to provide the volume overlap between the excitation light and the total volume of strips being excited, large enough to provide luminescence with sufficient intensity for the spectrometer to resolve. The excitation light source had a rectangular beam spot with an area comparable to the rectangular area filled with the SC-$Cu_2O$ strips. Shown in Fig. 3 are three PL spectra; The spectrum PL-NSC in red was collected from the NSC-CuO region. The two PL spectra, PL-SC(||) in blue and PL-SC(+), were collected from the strip region by placing the rectangular excitation beam spot parallel (||) and perpendicular (+), respectively to the strips in the strip region. In other words, the only difference between the PL-SC(||) and PL-SC(+) spectra is the areal size of the overlap between the rectangular excitation beam spot and the strip region. As expected, the PL-NSC spectrum confirms that no appreciable radiative recombination takes place in the NSC-CuO region while the two spectra, the PL-SC(||) and PL-SC(+) spectra, exhibiting six distinctive peaks indicate that complex radiative recombination dynamics are present in the SC-$Cu_2O$ strip region. All PL peak intensities are higher for the PL-SC(||) than those of the PL-SC(+) spectrum presumably because the net volume of crystallized $Cu_2O$ being optically excited in the PL measurement is much larger in the PL-SC(||) spectrum than in the PL-SC(+) spectrum. There are six narrow emission peaks, centered respectively about 1.37 eV, 1.62 eV, 2.05 eV, 2.17 eV, 2.29 eV, and 2.56 eV. The 2.17 eV emission is most likely to originate from the band edge recombination in SC-$Cu_2O$ and has been widely reported [21, 22, 23, 24]. The 2.17 eV emission, however, is weak because radiative recombination at the fundamental band edge is dipole forbidden [22]. The 2.05 eV emission may be attributed to the first excitonic transition (n=1) associated with the yellow series of $Cu_2O$ [15, 23]. The binding energy of the first or yellow excitonic series has been calculated to be ~150 meV [24], which is comparable to 160 meV – the difference in energy between the peak position of the band edge emission at 2.17 eV and the 2.05 eV emission seen in Fig. 3. This 2.05 eV emission, along with the emissions centered around 1.37 eV and 1.62 eV, may also be associated with an inter-band energy levels related to oxygen defects [23]. The origin of the 2.29 eV and 2.56 eV emissions may also be excitonic, possibly originating from the green and the blue excitonic series that have their highest energy transitions at 2.304 eV and 2.624 eV, respectively [15, 25]. The narrow linewidth of the emission peaks in the PL spectrum would suggest the involvement of radiative transitions between discrete and well-defined energy levels.

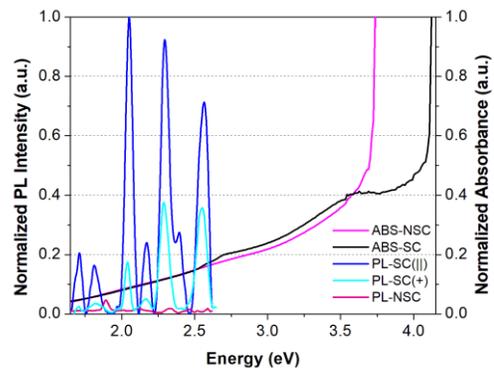

Fig. 3 Normalized photoluminescence and absorbance spectra of the NSC-CuO region and the SC-$Cu_2O$ strips.

The special coupon prepared for collecting the PL spectra was also used to obtain optical absorbance spectra shown in Fig. 3. The absorbance spectra, ABS-NSC (magenta) and ABS-SC (black), were collected from the NSC CuO region and SC $Cu_2O$ strip region, respectively, by measuring transmittance spectra using a Jasco v-670UV-Vis-NIR spectrophotometer in the energy range of 1.25 eV to 4.25 eV and subsequently converted to absorbance spectra. An appropriate shadow mask was used to discriminate the SC $Cu_2O$ strip region and the NSC region that coexist on the coupon. As seen in Fig. 3, the ABS-NSC spectrum (magenta) shows a monotonic increase with energy until it unveils a sharp increase reaching its maximum, approximately at 3.7 eV. The ABS-SC spectrum (black) shows an increasing trend similar to that of the ABS-NSC spectrum until it suddenly reaches its maximum, approximately 4.15 eV. There are three distinct features, in the ABS-SC spectrum, not seen in the ABS-NSC spectrum: a step-like feature at ~2.65 eV, another step-like feature at ~ 3.6 eV, and the energy at which the absorbance reaches its maximum is ~0.5 eV higher (i.e., a blue-shift) in comparison to that seen in the ABS-NSC spectrum. Amekura, H. et al. obtained absorption spectra of highly crystalline $Cu_2O$ nanostructures with dimensions < 50 nm, embedded in $SiO_2$, that showed step-like features, similar to those seen in Fig. 3, at ~2.65 eV and ~3.6 eV; the 2.65 eV step was ascribed to a transition in the blue exciton series and the step at 3.6 eV was attributed to a high energy X1-X3 transition in the Brillouin zone, additionally a marked absorption edge with a peak or step-like feature about the associated band edge energy of 2.17 eV is not seen because of the forbidden band edge transitions $Cu_2O$ is known for [18, 26]. If the marked features seen in the PL-SC and the ABS-SC spectra collected from the SC-$Cu_2O$ strips are associated with carrier dynamics involving excitons, they would need to originate from such environment as a quantum confined structures (e.g., quantum well) that strengthen the exciton binding energy, because all the measurements displayed in Fig. 3 were carried out at room temperature [27]. Since absorption spectra directly reflect the characteristics of the joint density of states in semiconducting materials, the presence of step-like features seen in the ABS-SC spectrum would suggest the presence of quantum confined structures in the SC- $Cu_2O$ strips [27]. Apart from the involvement of quantum confinement, the step-like features at ~2.65 eV and ~3.6 eV may suggest the involvement of inter-band transitions in the vicinity of the $\Gamma$–point often seen in high quality bulk $Cu_2O$, further indicating the presence of high crystallographic integrity of the SC-$Cu_2O$ strips [28, 29]. Detailed analytical transmission electron microscopy of the SC-$Cu_2O$ strips are currently underway and expected to correlate structural and chemical properties of the strips to the unique optical properties presented in Fig. 3.

## 4. Conclusion:

Laser-induced crystallization has been implemented for semiconductor thin films for decades; however its practical applications have been limited to only few successful demonstrations on thin films of single-element semiconductors and related devices exclusively designed to accommodate the major limiting factor of the laser crystallization: the use of femtosecond and excimer lasers. In this paper, we demonstrated the laser-induced crystallization of non-single-crystal CuO into single crystal $Cu_2O$, a multi-element semiconductor, using CW LD with a μ-CLB. The SC-$Cu_2O$ strips had a length extending to several millimeters and width reaching 10 μm. The optical studies done on the SC-$Cu_2O$ strips at room temperature revealed complex - unusual - emission and absorption characteristics most likely associated with excitonic transitions, suggesting the presence of quantum-confinement effects, this was not explicitly intended in our laser-induced crystallization process. Further assessment on the SC-$Cu_2O$ strips using such analytical tools as cross-sectional transmission electron microscopy and x-ray photo emission spectroscopy is needed to address the correlation between structural and chemical properties and the unique optical properties of the strips. Nevertheless, our demonstration would pave a way to obtain single-crystal thin films of alloy semiconductors with quality and dimensions required for a range of devices not currently feasible.